\documentclass{article}

\usepackage[letterpaper,margin=1.0in]{geometry}
\usepackage[utf8]{inputenc}  % allows input of unicode chars
\usepackage{array}
\usepackage{soul}  % highlighting
\usepackage{xcolor}  % highlighting
\usepackage{setspace}  % line spacing
\usepackage{fancyhdr}  % fancy header option
\usepackage[hidelinks]{hyperref}
\usepackage{enumitem}  % list parameters

\usepackage{graphicx}
\graphicspath{ {Figures/Main_text/} }
\usepackage{caption}  % customized captions
\usepackage{subcaption}  % allows sub-caption
\usepackage{makecell}  % table formatting

\usepackage[round]{natbib}   % omit 'round' option if you prefer square brackets
\usepackage{authblk}  % allows advanced author formatting

\usepackage{amsmath}  % Provides advanced math environments and tools.
\usepackage{amssymb} % Adds additional math symbols (like \mathbb, \leqslant, etc.).
\usepackage{mathtools} %Extends amsmath with more math tools (like \coloneqq, \adjustlimits, etc.).
\usepackage{mathrsfs}  %Provides \mathscr{} for script-style math fonts.

\newcommand{\beginsupplement}{%
        \fancyhead[L]{Supplemental Material}
        \setcounter{table}{0}
        \renewcommand{\thetable}{S\arabic{table}}%
        \setcounter{figure}{0}
        \renewcommand{\thefigure}{S\arabic{figure}}%
     }
\newcommand{\stopsupplement}{%
        \setcounter{table}{0}
        \renewcommand{\thetable}{\arabic{table}}%
        \setcounter{figure}{0}
        \renewcommand{\thefigure}{\arabic{figure}}%
     }

\title{\texttt{LinkedNN}: a neural model of linkage disequilibrium decay for recent effective population size inference} 

\author[1]{Chris C R Smith}
\affil[1]{Department of Biology, Indiana University}
\date{\today}

\begin{document}

\maketitle

%%%%%%%%%%%%%%%%%%%%%%%%%%%%%%%%%%%%%%%%%%%%%%%%
%%%%%%%%%% Abstract %%%%%%%%%%
%%%%%%%%%%%%%%%%%%%%%%%%%%%%%%%%%%%%%%%%%%%%%%%%
\section*{Abstract}
%Abstracts for Applications Notes are much shorter than those for an Original Paper. They are structured with four headings: Summary, Availability and Implementation, Contact and Supplementary Information.

\subsection*{Summary}
%Summary: This section should summarize the purpose/novel features of the program in one or two sentences.
A bioinformatics tool is presented for estimating recent effective population size that uses a neural network to automatically compute linkage disequilibrium-related features as a function of genomic distance between polymorphisms.
The new method outperforms existing deep learning and summary statistic-based approaches using relatively few sequenced individuals and variant sites, making it particularly valuable for molecular ecology applications with sparse, unphased data.

\subsection*{Availability and Implementation}
%Availability and Implementation: See above for advice and examples for this section.
The program is available as an easily installable Python package with documentation here: \sloppy{\url{https://pypi.org/project/linkedNN/}}.
The open-source code is available from: \sloppy{\url{https://github.com/the-smith-lab/LinkedNN}}.

\subsection*{Contact}
%Contact: Full email address to be supplied, preferably an institutional address.
Email: \url{chriscs@iu.edu}

\subsection*{Supplementary information}
%Supplementary information: Links to additional figures/data available on a web site, or reference to online-only Supplementary data available at the journal's web site.
Supplementary methods and figures are available via the journal's web site.

%%%%%%%%%%%%%%%%%%%%%%%%%%%%%%%%%%%%%%%%%%%%%%%%
%%%%%%%%%% Introduction %%%%%%%%%%
%%%%%%%%%%%%%%%%%%%%%%%%%%%%%%%%%%%%%%%%%%%%%%%%
\section*{Introduction}

Linkage disequilibrium (LD) is the non-random association of alleles at different loci and is shaped by evolutionary forces including recombination and genetic drift.
In particular, its magnitude reflects demographic history, where populations with larger effective size, $N_e$, show lower levels of LD \citep{hill1968linkage}.
This relationship has been used to develop powerful $N_e$ estimators that model correlations between genotypes at different loci as a function of genomic distance \citep{laurie1979allozymic, hill1981estimation, waples2006bias, santiago2020recent}.
LD-based approaches are especially useful for estimating \textit{recent} $N_e$, because crossover events occur more frequently than mutations and can therefore shape LD patterns over shorter timescales.
For example, \texttt{GONE} \citep{santiago2020recent} models the decay of LD with genomic distance to infer $N_e$ through time, producing reasonable $N_e$ estimates in the past 100 generations.
Such LD-based methods are especially valuable in molecular ecology and conservation where unphased and relatively sparse genomic markers can nonetheless provide signal about recent $N_e$.
Despite these advantages, summarizing genome-wide LD remains burdensome because existing approaches require analysis-specific decisions for binning SNP pairs into arbitrary, discrete distance classes.
Consequently, there is a need for tools that automatically extract features related to LD decay directly from polymorphism data.

Many of the deep learning architectures used to automatically extract features from SNPs are convolutional neural networks (CNNs). While CNNs have been effective for demographic inference and other tasks \citep{flagel2019unreasonable, torada2019imagene, gower2021detecting, sanchez2021deep, wang2021automatic, smith2023dispersal}, they have limited capacity to see LD.
This issue arises in part because CNNs perform spatial operations designed for gridded, regularly spaced inputs like images, but SNPs are not uniformly spaced along chromosomes.
Furthermore, the genotype information from individual sites is progressively eroded by layers beyond the first convolution, making typical architectures not ideal for seeing correlations beyond adjacent SNPs in the input array. 
Capturing precise LD signal requires models that can represent genomic features across a continuum of genomic distances.

Here, a neural network is developed that is capable of learning LD decay-related features from SNPs as a function of genomic distance, rather than relying on grid operations to convey positional information.
The method is evaluated on simulated data for estimating recent $N_e$, benchmarked against current CNN and summary statistic-based regression tools, and demonstrated on empirical data.
The new LD layer is implemented in a bioinformatics tool called \texttt{LinkedNN} that can be used for $N_e$ estimation or other inference tasks in diverse species.

%%%%%%%%%%%%%%%%%%%%%%%%%%%%%%%%%%%%%%%%%%%%%%%%
%%%%%%%%%% Methods %%%%%%%%%%
%%%%%%%%%%%%%%%%%%%%%%%%%%%%%%%%%%%%%%%%%%%%%%%%
\section*{Methods}

\subsection*{Neural network architecture}
The LD layer was implemented using PyTorch \citep{ansel2024pytorch} and consists of the following steps.
The initial stage involves sub-sampling pairs of SNPs, as the number of possible pairs scales quadratically with the input size.
Beginning with ordered polymorphisms $i=1,\dots,M$, discretized log-uniform index jumps $\Delta_i=\text{floor}(\text{log-U}(1,M))$ are drawn to form pairs $(i, i+\Delta_i)$, resulting in physical distances $d_i$ that are roughly log-uniform.
The number of sampled pairs is increased by repeating this process ten times, discarding pairs with $i+\Delta_i > M$, to yield a total of $P = 10M - n_\text{skip}$ pairs indexed by $p=1,\dots,P$.
For example, with $M=5,000$ and ten proposed pairs per SNP, the total retained pairs is around $P \approx 44,000$.
This strategy surveys a range of genomic distances without enumerating all pairs.

\begin{figure*}[!ht] % (Note the star symbol which makes the caption wide)
    \centering
    \includegraphics[width=1\textwidth]{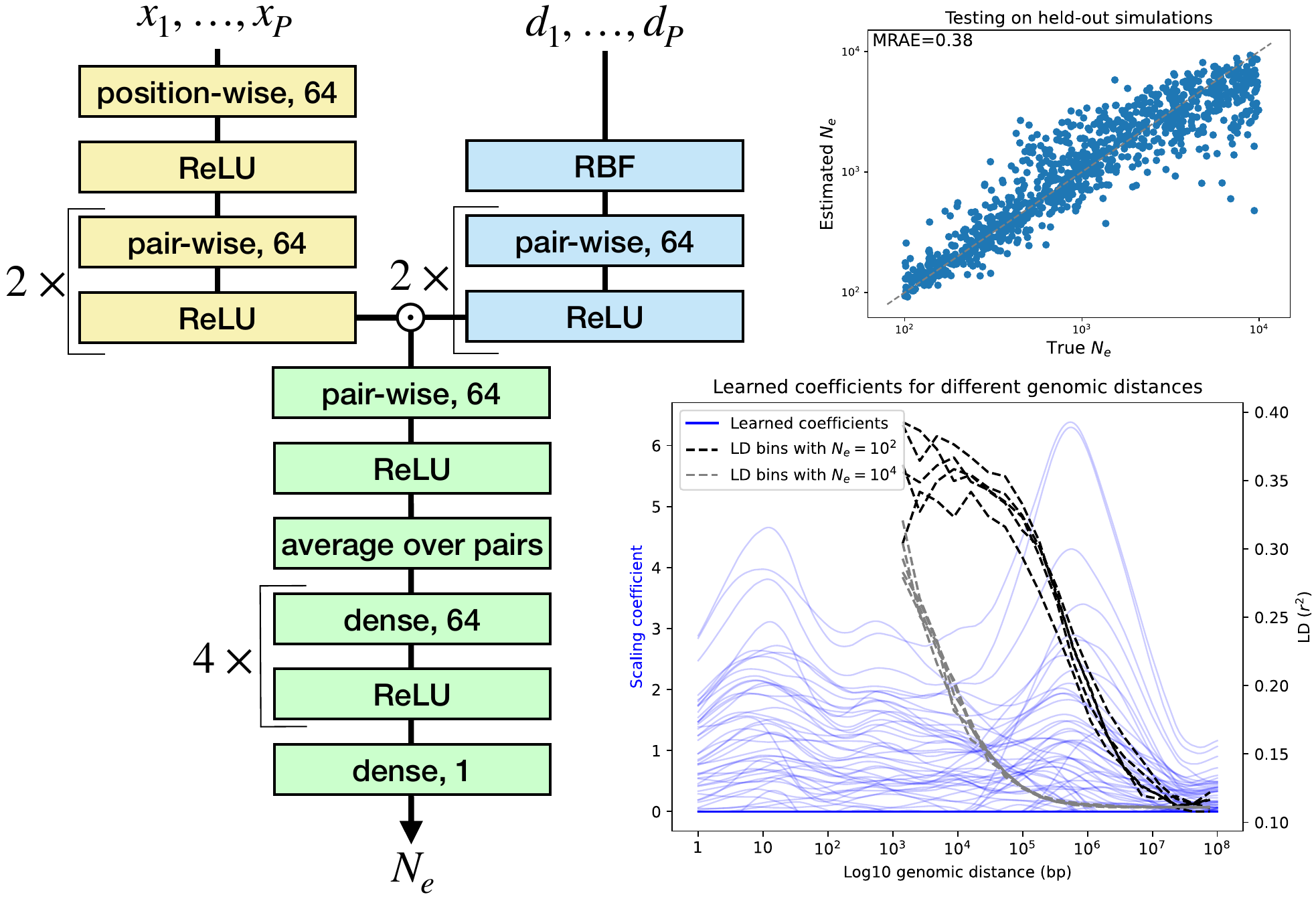}    
    \caption{
    \textit{(Left)} Neural network diagram. The inputs are genotypes for all SNP pairs $x_1, \dots, x_P$ and corresponding genomic distances $d_1, \dots, d_P$. 
    The number of filters is $f=64$ and rectified linear unit (ReLU) activation is used on all trainable layers except the final layer.
    Radial basis functions (RBF) are applied to the raw distances.
    The output can be effective population size, $N_e$, or another target the user defines.
    \textit{(Top right)}. Evaluating the LD layer on 1,000 simulations held out from training. 
    Axes are log-scale.
    The performance metric is mean relative absolute error (MRAE).
    \textit{(Bottom right)}. Blue lines are $f=64$ different coefficients output by the distance-mapping network of the pre-trained LD layer for a range of genomic distance inputs, $d_1, \dots, d_P$, irrespective of genotypes.
    Larger values indicate distances the model thinks are important for scaling particular genetic features.
    Black and grey dashed lines are binned $r^2$ values calculated on SNP pairs from five $N_e=10^2$ and five $N_e=10^4$ simulations, respectively.
    Bins for distances smaller than 1,000 bp were omitted because they contain too few SNPs.
    }
    \label{fig:fig1}
\end{figure*}

Features are initially extracted from the genotype inputs $x_1,...,x_P$ irrespective of genomic position (Figure \ref{fig:fig1}, left). 
Specifically, unphased genotypes are encoded as the count of the minor allele in each individual and given to a shared-weights, position-wise layer. 
All trainable layers have a number of output features $f=64$ and rectified linear unit activation, except the final layer.
The position-wise features in each pair are combined and given to two additional layers to compute preliminary genetic features $g_p$ for each pair.
By letting the network automatically extract features instead of manually calculating genotype correlations, it has the potential to extract non-LD features as well.

Next, the inter-SNP distances undergo further processing before being passed to any trainable layers.
The $d_p$ are transformed using radial basis functions, following the approach from \cite{schutt2017schnet} but applied in log space:
$$
e_k(d_{p})= \text{exp}(-\gamma ( \text{log}(d_{p}) - \mu_k )^2) 
$$
where the $K$ centers, $\mu_k$, are log-uniformly spaced over $(1,L)$, with $L$ being the length of the chromosome, and $K$ is set to $\text{log}(L)$, rounded up. 
The parameter $\gamma$ controls the width of the radial basis functions and, in this study, is implicitly set as the spacing between adjacent centers.
In effect, the feature space of the input is expanded from a single distance to a vector of features that activate depending on which center the input distance is closest to.
In the present study the radial basis functions are intended to smoothly partition distances into overlapping bins, allowing the neural network to diversify what information it uses for different distance ranges.
Base pair positions were used, as this approach works for non-model species lacking a genetic map, although centimorgan positions are superior and can be equivalently input to the LD layer, setting $L$ equal to the map length.

Whereas some CNNs have used genomic positions or distances as an additional channel in the input genotype array, this only permits the distance information to be conveyed additively to each filter, as the initial convolution performs a moving dot product between the input and kernel weights.
To allow the distances to interact with extracted genetic features multiplicatively, the LD layer learns a set of distance coefficients specific to each SNP pair, $s_p$, with length equal to the number of extracted genotype features, $f$.
The $s_p$ generating network consists of two layers applied to the radial basis features from each SNP pair.
These coefficients are used to directly rescale the corresponding genotype features from each SNP pair, elementwise: $g_p'= g_p \odot s_p$.
As a result, certain genotype features from some SNP pairs are amplified if their distance is relevant to the inference, or silenced if their distance is less pertinent.
This spatial mapping network addresses a similar need as the continuous filter convolution from \cite{schutt2017schnet} or edge-conditioned filters from \cite{simonovsky2017dynamic}, as the learned features are conditioned by a learned function of distance between data points.
However, instead of message-passing between nodes as with graph convolutions, the LD layer focuses on the interactions between nodes (SNP pairs), and uses fewer trainable parameters.
A similar approach was used by \cite{smith2024estimation} to rescale genotype features based on two-dimensional geographic distances between individuals.

After scaling by genomic distance, the $g_{m}'$ are each input to an additional layer, and then averaged across all SNP pairs. 
This averaging is inspired by the approach used by \cite{hill1981estimation} to combine data from all SNP pairs.
Last, the averaged features are input to a regression head consisting of five additional dense layers for estimating $N_e$.
The number of trainable weights in the LD layer is relatively small compared to a typical CNN, although training is slower due to computing many SNP pairs.
Training uses simulated datasets that match a focal empirical dataset in terms of sample size and other parameters.
In particular, a reference genome and recombination rate estimate or genetic map are required for the method.
Additional details are available in the Supplementary Material.

%%%%%%%%%%%%%%%%%%%%%%%%%%%%%%%%%%%%%%%%%%%%%%%%
%%%%%%%%%% Results %%%%%%%%%%
%%%%%%%%%%%%%%%%%%%%%%%%%%%%%%%%%%%%%%%%%%%%%%%%
\section*{Results}

\subsection*{Model evaluation}

The LD layer was compared with existing methods for estimating three parameters in a two-epoch demographic history, using 1,000 held-out simulations with $n=10$ sampled individuals and $M=5,000$ SNPs.
The best performing model was the LD layer, predicting the recent $N_e$ parameter with mean relative absolute error (MRAE) of 0.380 (Figure \ref{fig:fig1}, top right).
Second, the pairwise-CNN architecture from \cite{smith2023dispersenn2} gave MRAE=0.422 for the recent $N_e$ parameter.
Next, summary statistic-based regression using a neural network or random forest gave MRAE=0.429 and MRAE=0.456, respectively.
Finally, applying a basic CNN produced an MRAE of 0.511; relative to the basic CNN, the LD layer improved the MRAE by 25.6\%.

\subsection*{Interpreting learned distance-coefficients}
\begin{figure*}[!ht] % (Note the star symbol which makes the caption wide)
    \centering
    \includegraphics[width=0.5\textwidth]{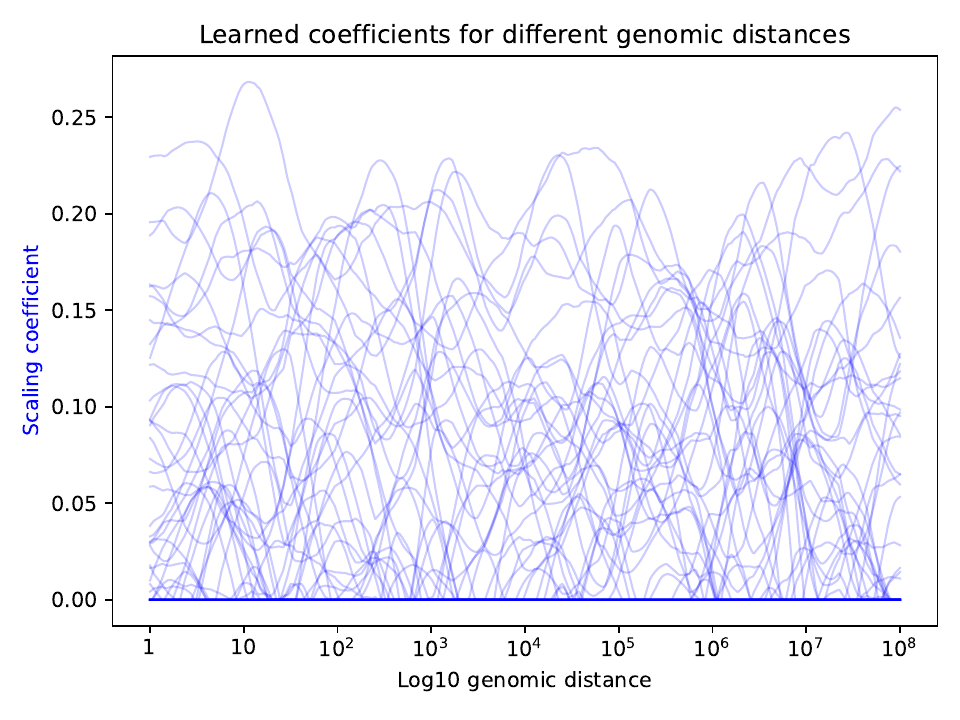}    
    \caption{
    Blue lines are $f=64$ different coefficients output by the distance-mapping network at initialization---without training---for a range of distance inputs.
    }
    \label{fig:positional_coefficients_atInit}
\end{figure*}

Next, the spatial coefficients, $s_p$, computed by the trained LD layer were examined for an array of input distances (Figure \ref{fig:fig1}, bottom right).
This inspection showed that the model learns unique coefficients corresponding to each genetic feature, with some coefficients displaying strong heterogeneity across the range of distances; this contrasts with the coefficients formed with random weight initializations which show no structure (Figure \ref{fig:positional_coefficients_atInit}).
In particular, 16 out of 64 coefficients had maximum values between $5 \times 10^5$ and $5 \times 10^6$ bp, indicating this was an important range of inter-SNP distances for the current analysis.
This distance is also larger than the mean spacing between consecutive SNPs in the input array of approximately $2 \times 10^4$, suggesting it is helpful to convey more distant SNP pairs than what the first layer of a CNN with kernel size of 2 would see.
In addition, the $5 \times 10^5$ to $5 \times 10^6$ bp sweet spot falls near the inflection point of the simulated LD-decay curve in small-$N_e$ simulations (Figure 1), suggesting the genetic features associated with these coefficients are related to LD.
Fifteen coefficients had maximum values between 50 and $5 \times 10^5$ bp, which may capture LD-related features at various stages of decay in larger populations, or correspond to genetic features unassociated with LD, e.g., measures of genetic variation.
From the remaining, nine features were canceled out by coefficients with flat values at zero, and 24 coefficients had maximum values between 5 and 50 bp; as SNP pairs rarely fell into the latter range, a likely explanation for these peaks is recognition of individual training examples, i.e., overfitting.
    
\subsection*{Empirical application}
Last, the LD layer was evaluated on publicly available data from harbor porpoises (\textit{Phocoena phocoena}) \citep{celemin2025evolutionary}, using 5,000 SNPs from the largest contig (maximum distance $6.7 \times 10^7$ bp) in $n=10$ genotypes from the Belt Sea group.
Uncertainty was assessed using 100 parametric bootstrap replicates, predicting on datasets simulated under the inferred demographic history.
The estimated recent $N_e$ under the two-epoch model was 1,411 (95\% CI: 745 to 3,493), with a population size change occurring 42.1 (31.5, 144.6) generations ago, and older $N_e$ of 5,921 (4,890, 7,665).
The inferred time parameter corresponds to 501.0 years in the past, applying a generation time of 11.9 years \citep{celemin2025evolutionary}.
These estimates seem reasonable, given the analyzed specimens were sampled geographically intermediate between the critically endangered Proper Baltic Sea sub-population \citep{carlstrom2023phocoena} and the Atlantic where larger populations exist \citep{celemin2025evolutionary}.
The inferred $N_e$ was smaller than that estimated by \cite{celemin2025evolutionary} using a sequentially Markovian coalescent-based method \citep{terhorst2017robust} which might struggle to estimate very recent demographic history.

%%%%%%%%%%%%%%%%%%%%%%%%%%%%%%%%%%%%%%%%%%%%%%%%
%%%%%%%%%% Discussion %%%%%%%%%%
%%%%%%%%%%%%%%%%%%%%%%%%%%%%%%%%%%%%%%%%%%%%%%%%
\section*{Discussion}

The new layer automatically computes LD-related features as a function of genomic distance between SNPs for estimating effective population size.
Evaluated on simulations, the LD layer was more accurate than machine learning-based regression models using mainstream summary statistics and CNN-based models.
While the inclusion of ideal summary statistics, such as carefully tuned LD-decay bins, might narrow this performance gap, the LD layer bypasses such an analysis by automatically learning which inter-SNP distances are useful.
The new method performs well with $n=10$ individuals and $5,000$ variants, making it valuable for molecular ecology applications with sparse, unphased SNPs, for example, from reduced representation sequencing.
However, the usefulness of LD-information will depend strongly on the analyzed demographic history and the number of recombination events between analyzed SNPs.
Future research may diversify the set of features extracted from SNPs by combining the LD layer with other architectures, which might help estimate complex demographic histories.

\bibliography{refs} %%% REFS %%%

 %%% SUPP MAT %%% 
\pagebreak

\beginsupplement

\section*{Supplementary material}

\subsubsection*{Model evaluation framework}
Simulations for training and evaluation were implemented in \texttt{msprime} with sample size of $n=10$ diploid individuals.
A two-epoch demographic scenario was used with ancestral size $N_a$ drawn from $\text{log-U}(10^2, 10^4)$, a population size change at time $T_1=\text{log-U}(10, 10^3)$ generations in the past, and recent size $N_1=\text{log-U}(10^2, 10^4)$.
Mutations were incrementally added to each simulated tree sequence until a fixed number of $M=5,000$ SNPs was achieved; this strategy prevents the model from using the number of segregating sites, which in many datasets reflects sequencing depth rather than evolutionary processes.
The vector including all three $z$-score-normalized parameters was used as the target for training several models including: (i) the LD layer, (ii) a basic CNN, (iii) a pairwise-CNN, and summary statistic-based regression with (iv) a random forest and (v) a neural network.
The size of the training set was 50,000 simulations---80\% for training, 20\% for hyperparameter tuning---and 1,000 simulations were held out for testing.

Training the LD layer and CNN-based models used batches of 100 training examples, mean squared error loss, the Adam optimizer, and starting learning rate of $1 \times 10^-4$.
The learning rate was halved every ten training iterations without reduction in validation loss.

\subsubsection*{Choice of CNN-based models}
The basic CNN from \cite{smith2023dispersal} was selected from among the several available CNN architectures because it is tuned for smaller SNP datasets, but this model is otherwise similar to that of \citep{flagel2019unreasonable}.
The pairwise-CNN is a relatively new architecture from \cite{smith2023dispersenn2} that was originally built for spatial applications but has not been benchmarked for estimating recent $N_e$.
Equivalent training settings---batch size, optimizer, etc.---were used with each CNN-based model for direct comparison with the LD layer.

\subsubsection*{Summary statistic-based regression}
The following summary statistics were calculated for each simulated dataset with \texttt{scikit-allel} \citep{miles2021cggh}:
\begin{itemize}[noitemsep]
    \item mean, across SNPs, of the average number of pairwise differences between chromosomes
    \item variance of the mean number of pairwise differences among SNPs 
    \item mean Tajima's D among non-overlapping 100-SNP windows
    \item variance of mean Tajima's D among windows
    \item mean observed rate of heterozygosity among SNPs
    \item variance of the observed rate of heterozygosity among SNPs 
    \item mean expected rate of heterozygosity among SNPs
    \item variance of the expected rate of heterozygosity among SNPs
    \item mean inbreeding coefficient among SNPs
    \item variance of the inbreeding coefficient among SNPs
    \item mean LD among pairs of variants calculated as $r$ from \cite{rogers2009linkage}
    \item variance in LD among pairs of variants
    \item each element of the folded allele frequency spectrum (ten bins for $n=10$ diploids)
\end{itemize}

Each summary statistic was $z$-score normalized (for the neural network). 
The random forest and neural network regressors from \texttt{scikit-learn} \citep{scikit-learn} were each trained with default settings on the same 50,000 simulations used to train the CNN-based and LD models, and tested on the same 1,000 held-out simulations.

\begin{table}[!ht]
\small
\centering
\begin{tabular}{||c l l l l||}
 \hline
  Parameter & Model  & RMSE & $r^2$ & MRAE \\ 
\hline\hline
    $N_1$ & CNN & 1587.8  & 0.795    & 0.511      \\      
    & summary statistics + random forest & 1480.2     & 0.838    & 0.456      \\
    & summary statistics + neural network & 1450.9     & 0.847    & 0.429      \\
    & pairwise-CNN & 1507.4  & 0.842    & 0.422      \\
    & \textbf{LD layer} & \textbf{1472.7}  & \textbf{0.86}     & \textbf{0.38}  \\
    
    $T_1$ & CNN & 166.5   & 0.581    & 0.749     \\
    & summary statistics + random forest & 191.4     & 0.561    & 0.760     \\
    & summary statistics + neural network & 193.8     & 0.567    & 0.687     \\
    & pairwise-CNN & 171.6   & 0.63     & 0.682     \\
    & \textbf{LD layer} & \textbf{150.5}   & \textbf{0.671}    & \textbf{0.578} \\
      
    $N_a$ & CNN & 611.2   & 0.972    & 0.172     \\
    & summary statistics + random forest & 1160.9     & 0.915    & 0.319     \\
    & summary statistics + neural network & 1154.9     & 0.920    & 0.314     \\
    & pairwise-CNN & 513.2   & 0.981    & 0.131     \\
    & \textbf{LD layer} & \textbf{485.3}   & \textbf{0.986}    & \textbf{0.121} \\
 \hline     
\end{tabular}
\caption{Model evaluation results. RMSE = root mean squared error, calculated on un-normalized outputs. The $r^2$ statistic was calculated on log-transformed outputs. MRAE = mean relative absolute error, calculated on un-normalized outputs.}
\label{table:results}
\end{table}

\stopsupplement

\end{document}